\documentclass[11pt]{article}
\usepackage{epsfig,epsf}
\usepackage{amsmath}
\usepackage{amsthm}
\usepackage{amsfonts}
\usepackage{amssymb}
\usepackage{psfrag}
\usepackage{cite}
\usepackage{slashed}
\usepackage{dsfont}
\usepackage{amscd}

\renewcommand{\theequation}{\arabic{section}.\arabic{equation}}

\renewcommand{\thetable}{\arabic{table}}

\renewcommand{\theequation}{\arabic{section}.\arabic{equation}}

\renewcommand{\thetable}{\arabic{table}}

\DeclareMathOperator{\sign}{sign}

\def \tr {\mathop{\rm tr}\nolimits}
\def \Im {\mathop{\rm Im}\nolimits}

\newcommand{\as}{\ifmmode\alpha_{\rm s}\else{$\alpha_{\rm s}$}\fi}
\newcommand{\asbar}{\ifmmode\bar{\alpha}_{\rm s}\else{$\bar{\alpha}_{\rm s}$}\fi}

\newcommand \CR {\mathcal{R}}

\newcommand{\bsigma}{{\boldsymbol{\sigma}}}

\newcommand{\brho}{{\boldsymbol{\rho}}}

\newcommand{\bel}{{\boldsymbol{\ell}}}

\font\cmss=cmss12 
\def\inbar{\,\vrule height1.5ex width.4pt depth0pt}
\def\IC{\relax\hbox{$\inbar\kern-.3em{\rm C}$}}
\def\IZ{\relax{\hbox{\cmss Z\kern-.4em Z}}}
\def\IR{{\hbox{{\rm I}\kern-.2em\hbox{\rm R}}}}

\def\IP{{\hbox{{\rm I}\kern-.2em\hbox{\rm P}}}}
\def\II{\hbox{{1}\kern-.25em\hbox{l}}}

\newbox\lett\newdimen\lheight\newdimen\lwidth
\def\ontop#1#2{\setbox\lett=\hbox{#2}\lheight\ht\lett
\multiply\lheight by 12 \divide\lheight by 10\relax%
\lwidth\wd\lett \multiply\lwidth by 8 \divide\lwidth by 10\relax #2\kern-\lwidth%
\raise\lheight\hbox{{$\scriptstyle #1$}}\kern.1ex}

\def\inbar{\,\vrule height1.5ex width.4pt depth0pt}

\numberwithin{equation}{section}

\begin{document}

\begin{titlepage}

\vspace*{20mm} \noindent
  {\Large \bf Noncompact $sl(N)$ spin chains: BGG~--~resolution, $\mathcal{Q}$-operators and
alternating sum representation for finite dimensional transfer matrices.

 }

\vspace{5mm} \noindent {\it \large  Sergey \' E. Derkachov~$^\dag$} and {\it \large
Alexander N. Manashov~$^{\ddag\S}$}
\\
\vspace*{0.1cm}

\noindent $^\dag$~{\it St.Petersburg Department of Steklov Mathematical Institute of
Russian Academy of Sciences, Fontanka 27, 191023 St.-Petersburg, Russia.}

E-mail:~derkach@pdmi.ras.ru \vskip 5mm

\noindent {$^\ddag$~\it Institute for Theoretical Physics, University of  Regensburg,
D-93040 Regensburg, Germany.\\
$^\S$~ Department of Theoretical Physics,  Sankt-Petersburg  University, St.-Petersburg,
Russia. }

 E-mail:~alexander.manashov@physik.uni-regensburg.de

\setcounter{footnote} 0

\noindent \vskip 1cm

\hskip 10mm
\begin{abstract}{
%{\small {\bf Abstract:}
 We study  properties of  transfer matrices
in the $sl(N)$ spin chain models. The transfer matrices with an infinite dimensional
auxiliary space are factorized into the product of $N$ commuting Baxter $\mathcal{Q}-$operators.
We consider the transfer matrices with auxiliary spaces of a special type (including the
finite dimensional ones). It is  shown that they can be represented as the alternating sum
over the transfer matrices with  infinite dimensional auxiliary spaces. We show that
certain combinations of the Baxter $\mathcal{Q}-$operators can be identified with the
$Q-$functions which appear in the Nested Bethe Ansatz. }
\end{abstract}

%\classification{Mathematics Subject Classification (2000)}{17B37,
%47N20, 81R50}
%
%\keywords{Yang-Baxter equation, Q-operator, BGG-resolution.}
%
%\end{opening}

%\vskip1cm
%
%
%
%\end{titlepage}

%{\tableofcontents}

%\newpage

%\end{titlepage}

%{\tableofcontents}

%\newpage
\vskip1cm

\end{titlepage}
%%%%%%%%%%%%%%%%%%%%%%%%%%%%%%%%%%%%%%%%%%%%%%%%%%%%%%%%%%%%%%%%%%%%%%%%%%%%%%%%%%%%%%%%%%%%%%
\section{Introduction}
%%%%%%%%%%%%%%%%%%%%%%%%%%%%%%%%%%%%%%%%%%%%%%%%%%%%%%%%%%%%%%%%%%%%%%%%%%%%%%%%%%%%%%%%%%%%%%
The $\mathcal{R}-$matrx approach~\cite{FST,KuSk1,Skl,Fad} to the
theory of integrable spin chain models allows one to construct an
infinite set of commuting operators -- transfer matrices which are
defined as the trace  of monodromy matrix over an auxiliary space
which can be both finite and infinite dimensional one. Transfer
matrices obey certain functional relations --- the so~-~called
fusion relations~\cite{KuSk1,Fusion,KR82} which provide one with a
powerful method to analyze  transfer matrices with finite
dimensional auxiliary space ($=$ finite dimensional transfer
matrices) (see e.g.
Refs.~\cite{Kuniba,Zabrodin,Tsuboi,KLWZ,Tsuboi98}). The derivation
of fusion relations is based heavily on the properties of  tensor
products  of representations. Namely, in much the same way as any
finite~-~dimensional representation (we restrict ourselves to the
case of $sl(N)$ algebra) can be obtained by tensoring the
fundamental representations any finite~-~dimensional transfer
matrix can be expressed in terms of the transfer matrices with
special representations in the auxiliary
space~\cite{Fusion,KR,Bazhanov:1989yk}.

The alternative approach is to start the analysis with consideration of transfer matrices
with  infinite dimensional auxiliary spaces of  a general type, which we will refer to as
the generic transfer matrices. All other transfer matrices, including finite-dimensional
ones, can be obtained from the generic transfer matrices by some reduction procedure. The
main advantage of this approach is a surprisingly simple structure of the generic transfer
matrices. Namely, it was shown in Ref.~\cite{slN} that  the generic $sl(N)$ transfer matrix
factorizes into the product of $N$ commuting operators~\footnote{This property is a direct
consequence of the factorization property of the $sl(N)$ invariant
$\mathcal{R}-$operator~\cite{SD,DM06,slN}.}
\begin{eqnarray}\label{tq}
\mathsf{T}_\brho(u)\sim\mathcal{Q}_1(u+\rho_1)\,
%\left(\mathcal{P}\tau^{{H}}\right)^{-1}\,
\mathcal{Q}_2(u+\rho_2)\,
\ldots
%  \left(\mathcal{P}\tau^{{H}}\right)^{-1}\,\,
\mathcal{Q}_N(u+\rho_N)\,.
\end{eqnarray}
The parameters $\{\rho_k\}$ define the representation of the $sl(N)$ algebra in the auxiliary
space (all notations will be discussed in sect.~\ref{notations}). The operators $\mathcal{Q}_k(u)$
form a commutative family and usually referred to as the Baxter $\mathcal{Q}-$operators. All
dependence of the generic transfer matrix on the representation in the auxiliary space
comes through  the shifts of the spectral parameters of the factorizing (Baxter)
operators. Thus, Eq.~(\ref{tq}) gives a complete description of  generic transfer matrices
in terms  of $N$ Baxter $\mathcal{Q}-$operators.

Representations of $sl(N)$ algebra are not exhausted by the finite dimensional and generic
infinite dimensional representations, however. For special values of the parameters $\{\rho_k\}$
the generic representation becomes reducible and contains invariant subspaces, which
can be both finite and infinite dimensional. In particular any finite dimensional representation is
realized as an invariant subspace of some generic representation.
In this work we study transfer matrices with an auxiliary space of
a special type and show that they can be expressed in terms of the generic transfer matrices.
In particular we obtain the determinant formula
for the finite~-~dimensional transfer matrices $t_{\brho}(u)$
\begin{eqnarray}\label{Qdet}
t_{\brho}(u)\sim\det
\left|\begin{array}{cccc} \mathcal{Q}_1(u+\rho_1)& \mathcal{Q}_1(u+\rho_2)&
\ldots&\mathcal{Q}_1(u+\rho_N)\\
\mathcal{Q}_2(u+\rho_1)& \mathcal{Q}_2(u+\rho_2)&
\ldots&\mathcal{Q}_2(u+\rho_N)\\
\vdots&\vdots&\ddots&\vdots\\
\mathcal{Q}_N(u+\rho_1)& \mathcal{Q}_N(u+\rho_2)&
\ldots&\mathcal{Q}_N(u+\rho_N)
\end{array}
\right|\,.
\end{eqnarray}
This formula  is a direct consequence of the factorized
representation~(\ref{tq}) and the Bernstein~-~Gelfand~-~Gelfand
resolution for the finite dimensional modules~\cite{BGG}. The
determinant formula~(\ref{Qdet}) is quite natural from many points
of view and, to our knowledge, was first proposed  in
Ref.~\cite{BLZ-III,Pronko:1999gh}.
It contains in a concise form a lot of information about the spin chain model.
In particular, it can be shown that the formula~(\ref{Qdet}) implies
the Nested Bethe Ansatz equations for the eigenstates of transfer matrices.

Equations~(\ref{tq}) and (\ref{Qdet}) show that both generic and finite dimensional
transfer matrices can be expressed in terms of Baxter $\mathcal{Q}-$operators. The Baxter
operators were  object of an intensive study in the last decade. The method of Baxter
operators allows to construct solutions for  models which do not possess the pseudovacuum
state and cannot be solved by the Algebraic Bethe Ansatz (ABA). The first construction of
such operator was given in the seminal paper of Baxter~\cite{Baxter:1972hz}. Later on
nontrivial examples of Baxter operators were found for a number of models. These are,
mostly, models with a rank one symmetry algebra. Common approach for  constructing Baxter
operators is based,  in this case, on solving the so-called $T-Q$
relation~\cite{BzSt90,GP92,Volkov,SDQ,Hikami,Pronko,SklyaninB,SL2C,DKM-I,RW,KMS,DKM-II,Korff04,Korff-c,Bytsko}.
Straightforward attempts to apply such approach for the models with higher rank symmetry
groups encounter certain difficulties. Another approach, the so-called $q-$oscillators
approach~\cite{BLZ,BLZ-II,BLZ-III,BHK,TsuboiB,Kojima,Bazhanov:2010ts}, provides a regular
method for constructing Baxter $\mathcal{Q}-$operators. In certain aspects it is close to the
approach based on the $\mathcal{R}-$matrix
factorization~\cite{SD,DMsl2,DMsl3,BDKM-21,DM06,slN}.

The paper is organized as follows: In section~\ref{notations} we explain notations and
discuss properties of $sl(N)$ modules and intertwining operators. In  section~\ref{TM} we
remind the definition of $sl(N)$ transfer matrices and Baxter $\mathcal{Q}-$operators and
formulate our results for the transfer matrices with auxiliary spaces of a special type. In
section~\ref{BGG} we derive the alternating sum representation (determinant formula) for
the  transfer matrices of this type. In section~\ref{sect:NBA} the relation of the
constructed transfer matrices  with Nested Bethe Ansatz is discussed. Some technical
details are collected in Appendix.

%%%%%%%%%%%%%%%%%%%%%%%%%%%%%%%%%%%%%%%%%%%%%%%%%%%%%%%%%%%%%%%%%%%%%%%%%%%%%%%%%%%%%%%%%%%%%
\section{$sl(N)$ modules}\label{notations}
%%%%%%%%%%%%%%%%%%%%%%%%%%%%%%%%%%%%%%%%%%%%%%%%%%%%%%%%%%%%%%%%%%%%%%%%%%%%%%%%%%%%%%%%%%%%%
Let $\mathbb{V}$ be a vector space of polynomials of $N(N-1)/2$ complex variables,
$z_{ki}$, $1\leq i<k\leq N$ of arbitrary degree
\begin{eqnarray}\label{Vm}
\mathbb{V}=\Big\{p(z_{21},z_{31},\ldots, z_{NN-1}),\quad
\deg(p)<\infty \Big \}\,.
\end{eqnarray}
We define the lower triangular matrix
\begin{eqnarray}\label{z}
{z}=\left(
\begin{array}{ccccc}1&0&0&\ldots&0\\
z_{21}&1&0&\ldots&0\\
z_{31}&z_{32}&1&\ldots&0\\
\vdots&\vdots&\vdots&\ddots&\vdots\\
z_{N1}&z_{N2}&\ldots&z_{NN-1}&1
\end{array}\right)
\end{eqnarray}
and adopt a shorthand notation $p(z)\equiv p(z_{21},z_{31},\ldots,
z_{NN-1})$.

Let $D_{mn}$, $m>n$ and $E_{ik}$ be the following differential operators on
$\mathbb{V}$:
\begin{eqnarray}\label{Dex}
D_{mn}
&=&\sum_{j=m}^N z_{jm}\,\frac{\partial\phantom{z_{j}}}{\partial z_{jn}}\,,
\nonumber\\
E_{ik}&=&-\sum_{m\leq n}z_{km}\,\Bigl(D_{nm}+\delta_{nm}\,\rho_m\Bigr)\,(z^{-1})_{ni}\,,
\end{eqnarray}
where $\rho_k$, $k=1,\ldots,N$ are complex numbers subjected to the constraint
\begin{eqnarray}\label{}
\rho_1+\rho_2+\ldots+\rho_N=N(N-1)/2\,.
\end{eqnarray}
Let ${e}_{ki}$ be the generators of the $sl(N)$ algebra,
$$
[{e}_{ik},{e}_{mn}]=\delta_{km} {e}_{in}-\delta_{in}{e}_{mk}\,.
$$
The homomorphism $\pi^\brho$
\begin{eqnarray}
\pi^\brho: {e}_{ik}\to E_{ik}\,,
\end{eqnarray}
where $\brho$ is the $N$ dimensional vector,
 $\brho\equiv(\rho_1,\cdots,\rho_N)$,
defines
a representation of the $sl(N)$ algebra on the space~$\mathbb{V}$ or provides
$\mathbb{V}$ with a structure of the
 $sl(N)$ module.  We will denote such a module by $\mathbb{V}_\brho$.

%%%%%%%%%%%%%%%%%%%%%%%%%%%%%%%%%%%%%%%%%%%%%%%%%%%%%%%%%%%%%%%%%%%%%%%%%%%%%%%%%%%%%%%%%
\subsection{Submodules and intertwining operators}\label{sln-red-m}
%%%%%%%%%%%%%%%%%%%%%%%%%%%%%%%%%%%%%%%%%%%%%%%%%%%%%%%%%%%%%%%%%%%%%%%%%%%%%%%%%%%%%%%%%
The representation $\pi^\brho$ (the module $\mathbb{V}_\brho$)
is {\it irreducible}  if none of the differences $\rho_{ik}=\rho_i-\rho_k$, $i<k$, is a positive
integer~\cite{Verma,BGG}.
Otherwise, the module $\mathbb{V}_\brho$ has invariant
subspaces. Some of them can be obtained as kernels of intertwining operators.

The operator
$D_{k+1k}$ raised to the power $\rho_{kk+1}$ intertwines
the generators corresponding to  two different sets of the parameters $\brho$, namely
\begin{eqnarray}\label{ED}
D_{k+1k}^{\rho_{kk+1}}\,E_{nm}(\brho)=E_{nm}(\brho')\,D_{k+1k}^{\rho_{kk+1}}\,,
\end{eqnarray}
where $\brho'$ differs from $\brho$ by permutation of
$\rho_k$ and $\rho_{k+1}$:
$\brho'=(\ldots,\rho_{k+1},\rho_k,\ldots)=P_{kk+1}\brho$.
Whenever $\rho_{kk+1}\in\mathbb{N}$, the operator
$D_{k+1k}^{\rho_{kk+1}}$ is a well defined operator on $\mathbb{V}$. It intertwines the representations
$\pi^\brho$ and $\pi^{\brho'}$ and its kernel
is an invariant subspace of $\mathbb{V}$.

We adopt a shorthand notation for the operators $D_{k+1,k}$,
${D}_k\equiv D_{k+1,k}$, $k=1,\ldots,N-1$. The operators commute,
${D}_k{D}_i={D}_i{D}_k$
if $|i-k|>1$,  while adjacent operators  satisfy
the following relations
\begin{eqnarray}\label{3D}
{D}_{k}^{a}\, {D}_{k+1}^{a+b}
\,{D}_{k}^{b}=
{D}_{k+1}^{b}\,{D}_{k}^{a+b}\,{D}_{k+1}^{a}\,.
\end{eqnarray}
These operators serve as elementary building blocks for
constructing  more general intertwining operators.
Let
$\brho_{ik}$ be a vector obtained from $\brho$ by a permutation of $i-$th and $k-$th elements,
$\brho_{ik}=P_{ik}\brho$.
The operator $U_{ik}$ which intertwines the generators
$E(\brho)$ and $E(\brho_{ik})$ has
the form (we assume that $i<k$)
\begin{eqnarray}\label{Example}
U_{ik}(\brho)=\left({{D}}_{k-1}^{\rho_{ik-1}}\ldots{{D}}_{i+1}^{\rho_{ii+1}}\right)
{{D}}_{i}^{\rho_{ik}}
\left({{D}}_{i+1}^{\rho_{i+1 k}}\ldots{{D}}_{k-1}^{\rho_{k-1 k}}\right)\,.
\end{eqnarray}
The statement that $U_{ik}$ is an intertwining operator  follows from~(\ref{ED}).
It is less obvious that  this operator
is a {\it polynomial} in $D_{nm}$ if  $\rho_{ik}=n>0$ and therefore
can be viewed as an operator on $\mathbb{V}$.
This can be easily checked with the help of commutation relations for the operators~$D_{mn}$
\begin{eqnarray}\label{}
[D_{mn},D_{ik}]=\delta_{in}D_{mk}-\delta_{mk} D_{in}\,.
\end{eqnarray}
Thus, whenever the difference $\rho_{i}-\rho_{k}$, $i<k$ is a positive integer
the operator $U_{ik}$ intertwines the representations $\pi^\brho$ and $\pi^{\brho_{ik}}$,
$\pi^{\brho_{ik}}U_{ik}=U_{ik}\pi^\brho$.

In a general situation (see Ref.~\cite{BGG}, Theorem 8.8), the operator $U$ which intertwines the
generators in the representations $\pi^\brho$ and $\pi^{\brho'}$,
is a  well defined operator on $\mathbb{V}$
 if and only if
\begin{itemize}
\item $\brho'$ can be
represented in the following form
$$\brho'=P_{i_n j_n}\ldots P_{i_1 j_1}\brho\,,\qquad (i_k<j_k)$$
\item
All differences $\rho^{(k)}_{i_{k+1}}-\rho^{(k)}_{j_{k+1}}=m_{k+1}\in \mathbb{N}$.
Here $\brho^{(k)}=P_{i_k j_k}\ldots P_{i_1 j_1}\brho$.
\end{itemize}
\vskip 5mm

Evidently, the kernel of an intertwining operator is  an invariant subspace of $\mathbb{V}$.
We will be interested in invariant subspaces of a special type.
Namely, let us consider a situation when the differences $\rho_{mm+1}$
are natural numbers starting from
$m=N-k$:
\begin{eqnarray}\label{rho-k}
\rho_{mm+1}=n_m\in \mathbb{N}, \qquad\mathrm{ for } \qquad m=N-k,\ldots N-1\,.
\end{eqnarray}
 In this case all operators ${D}_m^{\rho_{mm+1}}$ with
$m\geq  N-k$  have  nontrivial kernels.
Let us define an invariant submodule $\mathbb{V}^{(k)}_{\brho}$ as their intersection
\begin{eqnarray}\label{Vk}
\mathbb{V}^{(k)}_\brho= \ker {D}_{N-k}^{\rho_{N-kN-k+1}}\cap
\ldots\cap \ker {D}_{N-2}^{\rho_{N-2N-1}}
 \cap \ker{D}_{N-1}^{\rho_{NN-1}}\,.
\end{eqnarray}
The index $k$ shows the number of the intersecting spaces in~(\ref{Vk}). We will refer to
$\mathbb{V}^{(k)}_\brho$ as the space of the $k-$th level. Thus, a zero level space is the
generic space itself, $\mathbb{V}_{\brho}^{(0)}\equiv \mathbb{V}_{\brho}$, a space of the
first level $\mathbb{V}_{\brho}^{(1)}=\ker {D}_{N-1}^{\rho_{N-1N}}$ and so on. The
higher level of the space is the more restrictions  the functions from this space satisfy.
The space of the highest level for a given $N$, $\mathbb{V}^{(N-1)}_\brho$,
 is an irreducible finite dimensional $sl(N)$ module
(see Ref.~\cite{Zelobenko},
Chapter X).

 Taking into account that operator ${D}_k$
\begin{eqnarray}\label{exform}
{D}_{k}
=&\sum_{j=k+1}^N z_{jk+1}\,\frac{\partial\phantom{z_{j}}}{\partial z_{jk }}
\end{eqnarray}
 depends only on the elements in the
$k,k+1-$th columns of the matrix $z$
one finds
that for $k<N-1$ the subspace $\mathbb{V}^{(k)}_\brho$ can be represented as a tensor
product of an infinite dimensional and finite dimensional spaces,
\begin{eqnarray}\label{VslN}
\mathbb{V}^{(k)}_\brho={\mathbb{V}}_{k}\otimes {v}_{k}\,.
\end{eqnarray}
The elements of the space
${\mathbb{V}}_{k}$  are polynomials of arbitrary  degree which depend on the variables in
the first $N-k-1-$columns of the matrix $z$. Elements of the finite dimensional space
$v_k$  are polynomials which depend on the variables in the last $k+1$ columns of the
matrix $z$ and are nullified by the operators ${D}_m^{\rho_{mm+1}}$, $m\geq N-k$. The space $v_k$
can be considered as a finite~-~dimensional $sl(k+1)$ module
$v_k=v_{\rho_{N-k}\ldots\rho_N}$.
 We will be interested in the
transfer matrices with the higher level space $\mathbb{V}^{(k)}_\brho$ as an auxiliary space.

%%%%%%%%%%%%%%%%%%%%%%%%%%%%%%%%%%%%%%%%%%%%%%%%%%%%%%%%%%%%%%%%%%%%%%%%%%%%%%%%%%%%%%%%%%%%%%%%%%%%%%
\section{Transfer matrices}\label{TM}
%%%%%%%%%%%%%%%%%%%%%%%%%%%%%%%%%%%%%%%%%%%%%%%%%%%%%%%%%%%%%%%%%%%%%%%%%%%%%%%%%%%%%%%%%%%%%%%%%%%%%%
By a definition the $\mathcal{R}-$operator is a  solution of the Yang-Baxter equation (YBE)
\begin{eqnarray}\label{YBE}
\mathcal{R}_{12}(u)\mathcal{R}_{13}(u+v)\mathcal{R}_{23}(v)=
\mathcal{R}_{23}(v)\mathcal{R}_{13}(u+v)\mathcal{R}_{12}(u)\,.
\end{eqnarray}
It was shown in Ref.~\cite{slN} that the
$sl(N)$ invariant  $\mathcal{R}-$operator acting on the tensor product of two generic
$sl(N)$ modules
$\mathbb{V}_{\bsigma}\otimes\mathbb{V}_{\brho}$
($\bsigma=\{\sigma_1,\ldots,\sigma_N\}$, $\brho=\{\rho_1,\ldots,\rho_N\}$)
\begin{eqnarray}\label{Rop}
&&\mathcal{R}_{12}(u):\mathbb{V}_{\bsigma}\otimes\mathbb{V}_{\brho}\mapsto
\mathbb{V}_{\bsigma}\otimes\mathbb{V}_{\mathbf{\rho}}\,,\nonumber\\[0mm]
&&[E_{ik}^{(\bsigma)}+E_{ik}^{(\brho)},\mathcal{R}_{12}(u)]=0
\end{eqnarray}
can be represented in the factorized form~\cite{DM06,slN}
\begin{eqnarray}\label{RVV-factform}
\CR_{12}(u)=P_{12}\,\mathbb{R}^{(1)}_{12}(u-\sigma_1+\rho_1)\,
\mathbb{R}^{(2)}_{12}(u-\sigma_2+\rho_2)\, \ldots\,
\mathbb{R}^{(N)}_{12}(u-\sigma_N+\rho_N)\,.
%\nonumber\\
\end{eqnarray}
Here $P_{12}$ is the permutation operator and $\mathbb{R}_{12}^{(k)}(u)$ are the
factorizing operators.

The transfer matrix is defined as the  trace of the monodromy matrix
over the auxiliary
space. In the case of  an infinite dimensional auxiliary space one has to ensure
the convergence  of the trace. To this end we consider the modified $\mathcal{R}$
operator~\cite{slN}
\begin{eqnarray}\label{Rtau}
\mathcal{R}_{12}(u,\tau)\equiv
\mathcal{R}_{12}(u,\tau_1,\ldots,\tau_{N-1}) = \tau^{H}
\,\mathcal{R}_{12}(u)\, ,
\end{eqnarray}
where $\tau^{H}$ is a shorthand notation for the following operator acting on the second
space in the tensor product $\mathbb{V}_{\bsigma}\otimes \mathbb{V}_{\brho}$
\begin{eqnarray}\label{Rtau1}
\tau^{H}\equiv
\tau_1^{H_1}\,\tau_2^{H_2}\cdots\tau_{N-1}^{H_{N-1}} =
\prod_{p=1}^{N-1}\tau_p^{H_p}.
\end{eqnarray}
The operators $H_p$ are defined as
follows~\footnote{The definition of the operators $H_k$ adopted here is differ from that in~\cite{slN}
by a constant.
 It is done to have simple commutation relations between the $\mathcal{R}$ matrix and
 intertwining operators (see Eq.~(\ref{UikR}))
}
\begin{eqnarray}\label{H-p}
H_p=&\sum_{k=1}^p (E_{kk}%+\rho_k
+k-N)=%\sum_{m=p+1}^N
\sum_{k=1}^p\left(-\rho_k+\sum_{m=p+1}^N
z_{mk}\partial_{z_{mk}}\right)\,.
\end{eqnarray}
The operators $\mathcal{R}_{12}(u,\tau)$ obey  YBE
\begin{eqnarray}\label{YBp}
\mathcal{R}_{12}(u,\tau)\,\mathcal{R}_{13}(v,\tau)\,\mathcal{R}_{23}(v-u)=
\mathcal{R}_{23}(v-u)\,\mathcal{R}_{13}(v,\tau)\,\mathcal{R}_{12}(u,\tau)
\end{eqnarray}
and serve to construct the transfer matrix
\begin{eqnarray}
    \label{deft}
{\sf T}_{\brho}(u,\tau) =\tr_{\brho} \Big\{\CR_{10}(u,\tau)\ldots \CR_{L0}(u,\tau)\Big\}\,.
\end{eqnarray}
Here  the index $\brho$ specifies the representation, $\pi^{(0)}=\pi^{\brho}$, of the $sl(N)$
algebra on the auxiliary space. The trace on the rhs of Eq.~(\ref{deft}) exists for
$\tau<1$ ($\tau_k<1,k=1,\ldots,N-1$) and gives rise to a well-defined operator ${\sf
T}_{\brho}(u,\tau)$, (for details see Ref.~\cite{slN}). For the sake of simplicity we will
consider  homogeneous spin chains only, i.e. assume that the representations of the
$sl(N)$ algebra on the quantum space at each site are equivalent,
$\pi^{\bsigma_1}=\pi^{\bsigma_2}=\ldots=\pi^{\bsigma_L}\equiv \pi^{\bsigma}$. In this case
the transfer matrix~(\ref{deft}) is factorized into  the product of the Baxter
$\mathcal{Q}-$operators
\begin{eqnarray}\label{TQ}
\mathsf{T}_\brho(u,\tau)=\left(\mathcal{P}\tau^{{H}}\right)^{N-1}\mathcal{Q}_1(u+\rho_1,\tau)\,
%\left(\mathcal{P}\tau^{{H}}\right)^{-1}\,
\mathcal{Q}_2(u+\rho_2,\tau)\,
\ldots
 % \left(\mathcal{P}\tau^{{H}}\right)^{-1}\,\,
\mathcal{Q}_N(u+\rho_N,\tau)\,,
%\nonumber\\
\end{eqnarray}
%}
where the operator $H$ ($H=(H_1,\ldots,H_{N-1})$) acts on the quantum space of the model
$\mathbb{V}_q=\mathbb{V}_{\bsigma}\otimes \ldots\otimes \mathbb{V}_{\bsigma}$,
\begin{eqnarray}\label{Cartan}
H_k=H_k^{(1)}+\ldots+H_k^{(L)}
\end{eqnarray}
 and $\mathcal{P}$ is
the operator of cyclic permutations
\begin{eqnarray}
\mathcal{P}\, P(z_1,\ldots,z_L) = P(z_L,z_1,\ldots,z_{L-1}).
\end{eqnarray}
The definition of the operators $\mathcal{Q}_k(u,\tau)$ mimics the definition of transfer matrix
\begin{eqnarray}\label{defQ}
\mathcal{Q}_k(u+\sigma_k,\tau)= \tr_{0}
\Big\{{R}_{10}^{(k)}(u,\tau)\ldots
{R}_{L0}^{(k)}(u,\tau)\Big\}\,.
\end{eqnarray}
The operators ${R}_{i0}^{(k)}(u,\tau)$ are expressed in terms of the factorizing
operators~(\ref{RVV-factform}) as follows
\begin{eqnarray}
{R}^{(k)}_{i0}(u,\tau)=\tau^{H}\, P_{i0}\,
\mathbb{R}^{(k)}_{i0}(u)\,,
\end{eqnarray}
where the operator $H$ (see Eq.~(\ref{H-p})) acts on the auxiliary space. The Baxter
$\mathcal{Q}-$operators commute with each other and also with the operator of cyclic permutation
$\mathcal{P}$ and the operators $H_k$~(\ref{Cartan})
\begin{eqnarray}\label{}
[\mathcal{Q}_k(u,\tau),\mathcal{Q}_m(v,\tau)]=
[\mathcal{Q}_k(u,\tau),\mathcal{P}]=[\mathcal{Q}_k(u,\tau),H_k]=0\,.
\end{eqnarray}
They also satisfy   the simple normalization condition
\begin{eqnarray}\label{QPH}
\mathcal{Q}_k(\sigma_k,\tau)=\mathcal{P}\,\tau^{H}\,,
\end{eqnarray}
where $\sigma_k$ are the parameters of the representation in
the quantum space, $\bsigma=(\sigma_1,\ldots,\sigma_N)$.

Note that the factorized representation~(\ref{TQ}) for the transfer matrix holds
for arbitrary representation $\pi^\brho$ in the auxiliary space, whether it is
irreducible or not. In case that the representation $\pi^\brho$ is reducible
 the restriction of the $\mathcal{R}$ operator to the
invariant subspace $ {\mathbb{V}'}$ gives rise to a new solution of  the YBE,
${\mathcal{R}'}$:
$$
{\mathcal{R}'}:\mathbb{V}\otimes {\mathbb{V}'}\mapsto\mathbb{V}\otimes{\mathbb{V}'}\,.
$$
In this way one can obtain all $sl(N)$
invariant solutions of the YBE starting from the generic $\mathcal{R}$ operator.
For a reducible module $\mathbb{V}_\brho$ the $\mathcal{R}-$operator  has  the block triangular form,
\begin{eqnarray}\label{RRp}
\mathcal{R}=\left(\begin{array}{cc} \mathcal{R}' & \star\\
                               0&{\mathcal{R}''}\end{array}\right)\,,
\end{eqnarray}
where $\mathcal{R}''$ is an $\mathcal{R}-$operator on
$\mathbb{V}\otimes\mathbb{V}''$ and $\mathbb{V}''$ is the factor
space $\mathbb{V}''=\mathbb{V}_\brho/{\mathbb{V}'}$. Thus the
trace  over $\mathbb{V}_\brho$ in Eq.~(\ref{deft}) decays into the
traces over ${\mathbb{V}'}$ and $\mathbb{V}''$, so that one gets
\begin{eqnarray}\label{TTT}
\mathsf{T}_\brho(u,\tau)=\mathsf{T}'_\brho(u,\tau)+\mathsf{T}''_\brho(u,\tau)\,.
\end{eqnarray}
Studying   reducible representations  one can, in principle, express  transfer
matrices with  arbitrary $sl(N)$ submodule as the auxiliary space    in terms of the
generic transfer matrices,~$\mathsf{T}_\brho(u,\tau)$.
In a general situation the solution of the problem is not known, however for the transfer
matrices of a special type which will be considered in the next subsection  such expression exists.

%%%%%%%%%%%%%%%%%%%%%%%%%%%%%%%%%%%%%%%%%%%%%%%%%%%%%%%%%%%%%%%%%%%%%%%%%%%%%%%%%%%%%%%%%%%%%%%%
\subsection{Higher level transfer matrices}\label{sect:nongeneric}
%%%%%%%%%%%%%%%%%%%%%%%%%%%%%%%%%%%%%%%%%%%%%%%%%%%%%%%%%%%%%%%%%%%%%%%%%%%%%%%%%%%%%%%%%%%%%%%%
%
Let consider an $\mathcal{R}-$operator on the tensor product
$\mathbb{V}_\bsigma\otimes\mathbb{V}_{\brho}$
with $\brho$ satisfying conditions~(\ref{rho-k}). The generic module
$\mathbb{V}_\brho$ has an invariant submodule $\mathbb{V}_\brho^{(k)}$.
Let the operator $\mathcal{R}^{(k)}$ be a restriction of the $\mathcal{R}-$operator to the subspace
$\mathbb{V}_\bsigma\otimes\mathbb{V}_\brho^{(k)}$.
We define a  transfer matrix of $k-$th level, $\mathsf{T}^{(k)}_{\brho}$, as follows~\footnote{
We  remind that we consider  a homogeneous spin chain. This restriction is in no way a  principal one
but allows one to keep some expressions shorter.
}
\begin{eqnarray}\label{nongeneric}
\mathsf{T}^{(k)}_{\brho}(u,\tau)=\tr_{\mathbb{V}_{\brho}^{(k)}}\mathcal{R}_{10}^{(k)}(u,\tau)
\ldots\mathcal{R}_{L0}^{(k)}(u,\tau)\,.
\end{eqnarray}
We  also will use  a special notation for finite dimensional transfer matrices,
$t_\brho(u,\tau) = \mathsf{T}^{(N-1)}_{\brho}(u,\tau)$. For later convenience we will also
assume that the normalization of the  $\mathcal{R}-$ operator is chosen in such a way that
it has  simple commutation relations with the intertwining operators~\footnote{In
particular, the $SL(N,\mathbb{C})$ induced normalization introduced in~\cite{slN} gives
rise to Eq.~(\ref{UikR}).}
\begin{eqnarray}\label{UikR}
U\,\mathcal{R}_{\bsigma\brho}(u)=\mathcal{R}_{\bsigma\brho'}(u)\, U\,,
\end{eqnarray}
where $U$ intertwines  the representations $\pi^{\brho}$ and
$\pi^{\brho'}$. Existence of such normalization is provided by the universal
$\mathcal{R}-$operator  construction~\cite{Khoroshkin}. Indeed, the universal $\mathcal{R}-$operator
is  a function of  algebra generators only and hence
satisfies~(\ref{UikR}).

\vskip 5mm
\noindent
The higher level transfer matrices have the following properties:
\begin{itemize}
\item The  transfer matrix~(\ref{nongeneric}) of the $k-$th level can be represented in the
factorized form
\begin{eqnarray}\label{Wk}
\mathsf{T}^{(k)}_{\brho}(u,\tau)=(\mathcal{P}\tau^H)^{k+1-N}\prod_{j=1}^{N-k-1}\mathcal{Q}_j(u+\rho_j,\tau)
%\ldots\mathcal{Q}_{N-k-1}(u+\rho_{N-k-1},\tau)
\,\,W^{(k)}_{\brho_k}(u,\tau)\,,
\end{eqnarray}
where $\brho_k=(\rho_{N-k},\ldots,\rho_N)$.
The operator $W^{(k)}_{\brho_k}(u,\tau)$
is given by the trace of a special monodromy matrix over the auxiliary space
$\mathbb{V}_{\brho}^{(k)}$
\begin{eqnarray}\label{}
W^{(k)}_{\brho_k}(u,\tau)=\tr_{\mathbb{V}_{\brho}^{(k)}}\mathcal{R}_{10}^{(kN)}(u,\tau)\ldots
\mathcal{R}_{L0}^{(kN)}(u,\tau)\,,
\end{eqnarray}
where
\begin{eqnarray*}\label{}
\mathcal{R}_{j0}^{(kN)}(u,\tau)=\tau^{-H_0} P_{j0}\mathbb{R}^{(N-k)}_{j0}(u-\sigma_{N-k}+\rho_{N-k})\ldots
\mathbb{R}^{(N)}_{j0}(u-\sigma_N+\rho_N)\,.
\end{eqnarray*}
We remind that  $\bsigma=(\sigma_1,\ldots,\sigma_N)$ are the parameters of the
representation  in
the quantum space.
$\mathbb{R}^{(k)}_{j0}(u)$ are the factorizing operators~(\ref{RVV-factform})~\footnote{
The operator $\mathcal{R}_{j0}^{(kN)}(u,\tau)$ is defined on the tensor product
of two generic spaces $\mathbb{V}\otimes\mathbb{V}$ so that the permutation operator
is unambiguously defined, however it has invariant subspace $\mathbb{V}\otimes\mathbb{V}_{\brho}^{(k)}$ }.
The proof of~(\ref{Wk}) follows exactly  the proof of Theorem~2 in Ref.~\cite{slN} where
 details can be found.
\item The operators $W^{(k)}_{\brho_k}(u)$, $k=1,\ldots,N-1$ are finite in the limit $\tau\to 1$.

      The finiteness of $W^{(k)}_{\brho_k}(u,\tau=1)$ follows from the fact that the trace in~(\ref{Wk})
      involves only finite sums (see Ref.~\cite{slN}, Theorem~2)~ and therefore the $\tau-$regulator
      can be safely removed.

\item The transfer matrix of $k-$th level~(\ref{Wk})
      admits the following (alternating sum) representation in terms
      of the generic transfer matrices
\begin{eqnarray}\label{ResT}
\mathsf{T}^{(k)}_{\brho}(u,\tau)=\sum_{P\in P_{k+1}} (-1)^{\sign({P})} \mathsf{T}_{P\brho}(u,\tau)\,,
\end{eqnarray}
where   the sum runs over all permutations of the set %of $k+1-$elements
$(\rho_{N-k},\ldots,\rho_N)$,
i.e.
$$
P\brho=(\rho_1,\ldots,\rho_{{N-k-1}},\rho'_{N-k},\ldots,\rho'_{N})
$$
and $\sign(P)$ is the parity of the permutation. Making use of the factorized representation  for the
generic transfer matrix~(\ref{TQ}) one derives from Eq.~(\ref{ResT})
\begin{eqnarray}\label{detW}
W^{(k)}_{\brho_k}(u,\tau)=
(\mathcal{P} \tau^{H})^{-k}
\det
\left|\begin{array}{cccc} \mathcal{Q}_{N-k}(u+\rho_{N-k},\tau)
&\ldots&\mathcal{Q}_{N-k}(u+\rho_N,\tau)\\
\vdots%&\vdots
&\ddots&\vdots\\
\mathcal{Q}_N(u+\rho_{N-k},\tau)&
\ldots&\mathcal{Q}_N(u+\rho_N,\tau)
\end{array}
\right|\,.
\end{eqnarray}
We remind  that the parameters $\brho$ in the above expression satisfy the condition $\rho_{mm+1}=n_m\in
\mathbb{N}$, $m\geq N-k$.

\end{itemize}
Equation~(\ref{ResT}) follows from the Bernstein~-~Gelfand~-~Gelfand resolution for the
finite dimensional modules~\cite{BGG}.

%%%%%%%%%%%%%%%%%%%%%%%%%%%%%%%%%%%%%%%%%%%%%%%%%%%%%%%%%%%%%%%%%%%%%%%%%%%%%%%%%%%%%%%%%%%%%%%%%%
\section{BGG resolution}\label{BGG}
%%%%%%%%%%%%%%%%%%%%%%%%%%%%%%%%%%%%%%%%%%%%%%%%%%%%%%%%%%%%%%%%%%%%%%%%%%%%%%%%%%%%%%%%%%%%%%%%%%
We give here some details of Bernstein~-~Gelfand~-~Gelfand
construction for the $sl(N)$ algebra. Let $v_{\brho}$ be a finite dimensional module and
$\mathbb{V}_\brho$ is a generic $sl(N)$ module~(\ref{Vm}).
Any permutation $P$ can be represented as a composition of some number of permutations $P_{kk+1}$,
$k=1,\ldots, N-1$. The minimal number of such permutation is called the length of the permutation $P$,
$\ell(P)$.

Let  $\mathbb{V}_k$ be the direct sum of the spaces $\mathbb{V}_{P\brho}$,
where $\ell(P)=k$
\begin{eqnarray}\label{DVk}
\mathbb{V}_k=\sum_{P, \ell(P)=k}\oplus\mathbb{V}_{P\brho}\,.
\end{eqnarray}
That is $\mathbb{V}_0=\mathbb{V}_{\brho}$,
$\mathbb{V}_1=\mathbb{V}_{P_{12}\brho}\oplus\mathbb{V}_{P_{23}\brho}
\oplus\ldots\oplus\mathbb{V}_{P_{{N-1N}}\brho}$ and so on. The last space has index
$s=N(N-1)/2$, $\mathbb{V}_{s}=\mathbb{V}_{\rho_{N}\rho_{N-1}\ldots\rho_1}$.

The exact sequence constructed in Ref.~\cite{BGG} has the following form
\begin{eqnarray}\label{BGGs}
\begin{CD}
0@>>>v_\brho @>\varepsilon>>\mathbb{V}_0@>{d_1}>>\mathbb{V}_1 @>d_2>>
\ldots
@>d_{s}>>
\mathbb{V}_{s}@>>>0
\end{CD}
\end{eqnarray}
Here the mapping $\varepsilon$ is the natural embedding of the module $v_\brho$ into
$\mathbb{V}_0$. The operator $d_k$ is the $n_{k}\times n_{k-1}$ matrix, where $n_k$ is the
number of the permutation of the length~$k$ (the number of the spaces in the direct
sum~(\ref{DVk})). Let $\brho^i$ and $\brho^j$ be the parameters of the generic modules
$[\mathbb{V}_k]_i$ and $[\mathbb{V}_{k-1}]_j$, respectively, (i.e.
$[\mathbb{V}_k]_i=\mathbb{V}_{\brho^i}$, $[\mathbb{V}_{k-1}]_j=\mathbb{V}_{\brho^j}$).
The entries $[d_k]_{ij}$ are nonzero only if $\brho^i$ and $\brho^j$  differ by permutation
of two elements, $\brho^i=P_{k n}\brho^j$ and
$(\rho^j)_k-(\rho^j)_n=m>0$ ($k<n$).
In this case
$[d_k]_{ij}=a^{k}_{ij} U_{kn}(\brho^j)$, where $U_{kn}$ is the intertwining operator~(\ref{Example}) and
the coefficients $a^{k}_{ij}$ are plus or minus one.
For instance, the first operator can be chosen as
$[d_1]_{i1}=D_{i}^{\rho_{ii+1}}$, $i=1,\ldots, N-1$.

Let us now fix the indices $i$ and $m$ and consider the matrix elements $[d_{k}]_{ij}$ and
$[d_{k-1}]_{jm}$.
It was shown in Ref.~\cite{BGG} that either all these elements are zero for all $j$ or
there exist exactly two numbers $j_1$ and $j_2$ for which $[d_{k}]_{ij}$ and
$[d_{k}]_{jm}$ are nonzero. Moreover, the numbers $a^{k}_{ij}=\pm1$ can be chosen in
such a way that the product $a^{k}_{ij_1}a^{k}_{ij_2}a^{k-1}_{j_1m}a^{k-1}_{j_2m}=-1$. This property
guarantees  that $[d_{k}]_{ij_1}[d_{k-1}]_{j_1m}+[d_{k}]_{ij_2}[d_{k-1}]_{j_2m}=0$ and
hence $d_k d_{k-1}=0$, i.e. $\Im d_{k-1}\subset \ker d_{k}$.

The sequence~(\ref{BGGs}) is called a (BGG) resolution of the finite~-~dimensional module.
It was proved by
Bernstein, Gelfand and Gelfand  that it is exact sequence,
i.e. $\ker d_{k}=\Im d_{k-1}$ for all $k$.
\vskip 5mm

Let $\mathbf{T}_k(u,\tau)$ be a monodromy matrix on the $\mathbb{V}_q\otimes\mathbb{V}_k$,
where $\mathbb{V}_q$ is the quantum space of the spin chain model and $\mathbb{V}_k$ is
the space~(\ref{DVk}).
The monodromy matrix $\mathbf{T}_k(u)$~\footnote{For brevity, till the end of this section
we will omit the $\tau-$dependence.
}
 is a diagonal $n_k\times n_k$ matrix. Its entries
are ordinary monodromy matrices, $[\mathbf{T}_k(u)]_{ii}=\mathsf{T}^{(i)}_k(u)$,
which are defined on the tensor product $\mathbb{V}_q\otimes[\mathbb{V}_k]_i$.
Obviously, the trace of  monodromy matrix
$\mathbf{T}_k(u)$ over
the space $\mathbb{V}_k$ is given by the sum of transfer matrices~$\mathsf{T}_{P_i\brho}(u)$
(see Eq.~(\ref{deft}))
\begin{eqnarray}\label{TtT}
\tr_{\mathbb{V}_k}\mathbf{T}_k(u)=\sum_{i=1}^{n_k}  \mathsf{T}_{P_i\brho}(u)\,.
\end{eqnarray}
Next, since $\ker d_{k}$ is the invariant subspace of $\mathbb{V}_{k-1}$ the trace decays
into the traces over $\ker d_{k}$ and the factor space
$\widetilde{\mathbb{V}}_{k-1}=\mathbb{V}_{k-1}/\ker
d_k=\Im d_k$. So long as $d_k\,\mathbf{T}_{k-1}=\mathbf{T}_{k}\,d_k$ one obtains
$$
\tr_{\widetilde{\mathbb{V}}_{k-1}}\mathbf{T}_{k-1}(u)=
\tr_{\Im d_k}\mathbf{T}_{k}(u)=\tr_{\ker d_{k+1}}\mathbf{T}_{k}(u)\,
$$
and hence
\begin{eqnarray}\label{kk}
\tr_{\mathbb{V}_{k-1}}\mathbf{T}_{k-1}(u)=\tr_{\ker d_k}\mathbf{T}_{k-1}(u)+
\tr_{\ker d_{k+1}}\mathbf{T}_{k}(u)\,.
\end{eqnarray}
Taking into account the boundary conditions
\begin{eqnarray*}
t_\brho(u)=\tr_{\ker d_1}\mathbf{T}_{0}(u)\,,&\phantom{hhhhh}&
\tr_{\ker d_{s+1}}\mathbf{T}_{s}(u)=\tr_{\mathbb{V}_{s}}\mathbf{T}_{s}(u)
\end{eqnarray*}
one derives from~Eqs.~(\ref{TtT}),~(\ref{kk})
\begin{eqnarray}\label{}
t_\brho(u)=\mathsf{T}_\brho^{(N-1)}=
\sum_{k=0}^{s}(-1)^k\mathbf{T}_{k}(u)=\sum_{P}(-1)^{\sign(P)} \mathsf{T}_{P\brho}(u)\,,
\end{eqnarray}
where the last sum runs over all permutations.
\vskip 3mm

Let us notice that the derivation of the representation~(\ref{ResT}) for the higher level transfer matrices,
$\mathsf{T}_\brho^{(k)}(u)$, $k<N-1$ follows exactly the same lines.
It is sufficient to notice that the submodule $\mathbb{V}^{(k)}_\brho$ is given by the tensor
product~(\ref{VslN}) where the space $v_k$ is a finite dimensional $sl(k+1)$ module.

%%%%%%%%%%%%%%%%%%%%%%%%%%%%%%%%%%%%%%%%%%%%%%%%%%%%%%%%%%%%%%%%%%%%%%%%%%%%%%%%%%%%%%%%%%%%%%%%
\section{Baxter Equation and Nested Bethe Ansatz}\label{sect:NBA}
%%%%%%%%%%%%%%%%%%%%%%%%%%%%%%%%%%%%%%%%%%%%%%%%%%%%%%%%%%%%%%%%%%%%%%%%%%%%%%%%%%%%%%%%%%%%%%%%
The finite dimensional module $v_\brho$,
$\rho_{kk+1}=n_k=m_k+1\in \mathbb{N}$ corresponds to the highest
weight $\chi=(m_1,\ldots,m_{N-1})$ or to the  Young tableau
specified by the partitions
$${\bel}=\{\ell_1,\ell_2,\ldots,\ell_{N-1}\},$$
 where
$\ell_k=\sum_{i=k}^{N-1} m_i$ is the length of the $k-$th row in
the tableau.
The finite dimensional transfer matrix corresponding to the Young
tableau $\bel$ has the form
\begin{eqnarray}\label{tell}
t_{\bel}(u+f_\bel,\tau)=(\mathcal{P} \tau^{H})^{-N+1}
\det
\left|\begin{array}{cccc} \mathcal{Q}_1(u+l_1,\tau)& \mathcal{Q}_1(u+l_2,\tau)&
\ldots&\mathcal{Q}_1(u,\tau)\\
\mathcal{Q}_2(u+l_1,\tau)& \mathcal{Q}_2(u+l_2,\tau)&
\ldots&\mathcal{Q}_2(u,\tau)\\
\vdots&\vdots&\ddots&\vdots\\
\mathcal{Q}_N(u+l_1,\tau)& \mathcal{Q}_N(u+l_2,\tau)&
\ldots&\mathcal{Q}_N(u,\tau)
\end{array}
\right|\,,
\end{eqnarray}
where $l_k=\rho_k-\rho_N=\ell_k-k+N$ and
$$
f_\bel\equiv-\rho_N=\frac1N\sum_{k=1}^{N-1}\ell_k=\frac1N(m_1+2m_2+\ldots+(N-1)m_{N-1}).
$$

The transfer matrix corresponding to the {\it null} Young tableau, $\ell_k=0$,
$k=1,\ldots,N-1$ (i.e. to the trivial one dimensional representation),
$t_{\bel=0}(u,\tau)$, is proportional to the unit operator on the quantum space,
\begin{eqnarray}\label{nullT}
t_{\bel=0}(u,\tau)=\Delta(u,\tau)\II.
\end{eqnarray}
 Thus, it follows
from~(\ref{tell}) that the determinant of a $N\times N$ matrix $A(u,\tau)$,
$A_{ij}(u,\tau)=\mathcal{Q}_{i}(u+N-j,\tau)$, is given by
\begin{eqnarray}\label{wronskian}
\det ||A(u,\tau)||=\Delta(u,\tau)\, (\mathcal{P}\tau^{H})^{N-1}\,.
\end{eqnarray}
The equation~(\ref{wronskian}) is usually referred to as the Wronskian relation.

\vskip 5mm The determinant representation~(\ref{tell}) gives rise to  the self-consistency
equations (Baxter equations) involving the Baxter $\mathcal{Q}-$operators and the finite
dimensional transfer matrices. Indeed, let $t_k(u)$ be the transfer matrix corresponding
to the Young tableau with one column and $k$~boxes, $0\leq k\leq N$, (the transfer
matrices with $k=0$ or $k=N$ correspond to the trivial representation, i.e.
$t_0(u,\tau)=t_N(u,\tau)=\Delta(u,\tau)$.)
Let $B_j(u)$ be  a $(N+1)\times (N+1)$ matrix
\begin{eqnarray}\label{}
(B_j)_{ik}(u,\tau)=\left\{\begin{array}{cc}\mathcal{Q}_i(u+N+1-k,\tau),& i\leq N\,,\\
\mathcal{Q}_j(u+N+1-k,\tau), &i=N+1\,\end{array}\right.
\,.
\end{eqnarray}
By construction the matrix $B_j$ has two identical lines, hence $\det B_j(u,\tau)=0$.
Expansion of  $\det B_j(u,\tau)$ over the elements of the last line
gives rise to the following relation
\begin{eqnarray}\label{BE}
\sum_{k=0}^{N}(-1)^k\, t_k(u+k/N,\tau)\, \mathcal{Q}_j(u+N-k,\tau)=0\,,
\end{eqnarray}
which is the $N-$th order difference (Baxter) equation on the operator $\mathcal{Q}_j(u,\tau)$.
Let us notice that except the $sl(2)$ case the Baxter operators $\mathcal{Q}_k(u,\tau)$ with $k<N$
are singular in the limit  $\tau\to1$. However, as follows from
Eqs.~(\ref{detW}),~(\ref{tell}) their antisymmetrized products are free from
singularities. \vskip 5mm

To establish connection with the Nested Bethe Ansatz approach let us consider the minors
of the matrix corresponding to the null Young tableau. Namely, we define new set of the
operators $\widehat{\mathcal{Q}}_k(u,\tau)$, $k=2,\ldots,N$ by
\begin{eqnarray}\label{}
\widehat{\mathcal{Q}}_{k}(u,\tau)=\det
\left|\begin{array}{cccc} \mathcal{Q}_{k}(u+N-1,\tau)& %\mathcal{Q}_1(u+N-k-1,\tau)&
\ldots&\mathcal{Q}_{k}(u+k-1,\tau)\\
\vdots%&\vdots
&
\ddots&\vdots\\
\mathcal{Q}_N(u+N-1,\tau)& %\mathcal{Q}_N(u+N-k-1,\tau)&
\ldots&\mathcal{Q}_N(u+k-1,\tau)
\end{array}
\right|\,,
\end{eqnarray}
that is
\begin{eqnarray*}
\widehat{\mathcal{Q}}_{N}(u,\tau)&=&\mathcal{Q}_N(u+N-1,\tau),\\
\widehat{\mathcal{Q}}_{N-1}(u,\tau)&=&\mathcal{Q}_{N-1}(u+N-1,\tau)\mathcal{Q}_{N}(u+N-2,\tau)%\nonumber\\
%&&
-
\mathcal{Q}_{N}(u+N-1,\tau)\mathcal{Q}_{N-1}(u+N-2,\tau)
\end{eqnarray*}
 and so on.
The operators $\widehat{\mathcal{Q}}_k$ arise as the factorizing operators for the $k-$th level
transfer matrices, such that $\rho_{mm+1}=1$, for $m\geq N-k$,
\begin{eqnarray}\label{}
\mathsf{T}_{\brho}^{(k)}(u,\tau)\sim \mathcal{Q}_1(u+\rho_{1},\tau)\ldots
\mathcal{Q}_{N-k-1}(u+\rho_{N-k-1},\tau)\,
\widehat{\mathcal{Q}}_{N-k}(u+\delta_k,\tau)\,,
\end{eqnarray}
where $\delta_k=\rho_N+k+1-N$.
Hence, as it was explained in~Sec.~\ref{sect:nongeneric}
these operators are finite in the limit
$\tau\to 1$.
\vskip 3mm

Further, with the help of the identity~(\ref{Ai}) one can express the ratio of transfer
matrices
$t_{1}(u+1/N)/t_0(u)\equiv \bar t_1(u)$ as follows
(henceforth we will omit $\tau-$dependence),
\begin{eqnarray}\label{t1Q}
\bar t_1(u)=
\frac{\Delta(u+1)}{\Delta(u)}\frac{\widehat{\mathcal{Q}}_{2}(u-1)}{\widehat{\mathcal{Q}}_{2}(u)}+
\sum_{k=2}^{N-1}\frac{\widehat{\mathcal{Q}}_{k}(u+1)}{\widehat{\mathcal{Q}}_{k}(u)}
\frac{\widehat{\mathcal{Q}}_{k+1}(u-1)}{\widehat{\mathcal{Q}}_{k+1}(u)}+
\frac{\widehat{\mathcal{Q}}_{N}(u+1)}{\widehat{\mathcal{Q}}_{N}(u)}
\,,
\end{eqnarray}
where $\Delta(u)=t_0(u)$.
Introducing the notations
\begin{eqnarray}\label{Lambdak}
\Lambda_k(u)
=\frac{\widehat{\mathcal{Q}}_{k}(u+1)}{\widehat{\mathcal{Q}}_{k}(u)}
\frac{\widehat{\mathcal{Q}}_{k+1}(u-1)}{\widehat{\mathcal{Q}}_{k+1}(u)}\,,
\hskip 1cm
k=1,\ldots,N,
\end{eqnarray}
where $ \widehat{\mathcal{Q}}_1=\Delta(u)$ and $\widehat{\mathcal{Q}}_{N+1}(u)=1$ one can
rewrite~(\ref{t1Q}) in the form
\begin{eqnarray}\label{NBE}
\bar t_1(u)=%
\Lambda_1(u)+\Lambda_1(u)+\ldots+\Lambda_N(u)\,,
\end{eqnarray}
which is the well-known expression
for the transfer matrix in the Nested Bethe Ansatz approach~\cite{Sutherland,Kulish83}.

Making use of the identity~(\ref{Ai}) and the determinant representation~(\ref{tell})
one can express the transfer matrices $\bar t_k(u)=t_k(u+k/N)/\Delta(u)$
corresponding to the Young tableau with one column and $k$ boxes
in
terms of ratios~(\ref{Lambdak})
\begin{eqnarray}\label{btk}
\bar t_{k}(u)=\sum_{m_1=1}^N\sum_{m_2=m_1+1}^{N}\ldots\sum_{m_k=m_{k-1}+1}^N
 \Lambda_{m_1}(u)\Lambda_{m_2}(u-1)\ldots\Lambda_{m_k}(u-k+1)\,.
\end{eqnarray}
The representation for an arbitrary finite dimensional transfer matrix in terms of
$\Lambda_k$ were obtained by Kirillov and Reshetikhin, see Ref.~\cite{KR,Bazhanov:1989yk} for details.

\vskip 3mm

The finite dimensional transfer matrices satisfy an infinite set of the functional
(fusion) relations~\cite{Fusion,Kuniba,Tsuboi,Tsuboi98}.
The representation~(\ref{tell}) allows one  to generate such
relations in a straightforward way with the help of the following determinant identity
\begin{eqnarray}\label{ab}
\|a_1,a_2,a_3,\ldots,a_N\| \|b_1,b_2,a_3,\ldots,a_N\|&=&
\|a_1,b_1,a_3,\ldots,a_N\| \|a_2,b_2,a_3,\ldots,a_N\|\nonumber\\
&&+
\|a_1,b_2,a_3,\ldots,a_N\| \|b_1,a_2,a_3,\ldots,a_N\|\,.\nonumber\\
\end{eqnarray}
Here $\|a_1,a_2,a_3,\ldots,a_N\|$ stands for the determinant of $N\times N$ matrix with
the columns $a_1,\ldots a_N$, and similar for others.
For example, let us assume that the Young tableau $\bel$ has more than one column and that
the last row with more than one box in the row has index $p$, i.e.  $\ell_p>1$ and
$\ell_{p+1}\leq 1$. Using Eq.~(\ref{ab}) one can derive the following quadratic relation
\begin{eqnarray}\label{fr-1}
\tilde t_{\bel}(u)\tilde t_{\bel'_1}(u+1)=\tilde t_{\bel'}(u+1)\tilde t_{\bel_1}(u)-
\tilde t_{\bel_{-}}(u)\tilde t_{\bel_+}(u+\delta)\,,
\end{eqnarray}
where we put $\tilde t_{\bel}(u)=t_{\bel}(u+f_{\bel})$. The Young tableau $\bel_1$ is
obtained from the tableau $\bel$ by deleting one box from the $p-$th row. The primed Young
tableau $\bel'(\bel'_1)$ is obtained from the corresponding unprimed tableau
$\bel(\bel_1)$ by crossing out the first column.  The Young tableau $\bel_-$ is obtained
from the tableau $\bel$ by crossing out all rows except first $p-1$, i.e.
$\bel_-=\{\ell_1,\ldots,\ell_{p-1},0\ldots,0\}$. To construct the table $\bel_+$ one has
to  cross out all columns containing $N-$boxes from the auxiliary table with $N$ rows
$$
L=\{\ell_1-1,\ldots\ell_{p-1}-1,\ell_p-1,\ell_p-1,\ell_{p+1},\ldots,\ell_{N-1}\}\,.
$$
 The parameter $\delta=1+L_N$,
i.e. $\delta=1$ for $\ell_{N-1}=0$, $\delta=2$ for $\ell_{N-1}=1$ and $\delta=\ell_p$ if
$p=N-1$. Let us introduce ordering for Young tableaux as follows: $\bel'<\bel$ if the
first nonzero entry in $\bel-\bel'=\{\ell_1-\ell'_1,\ell_2-\ell'_2,\ldots\}$ is positive.
It is easy to see that the Young tableau $\bel$ is the maximal one among the Young
tableaux appearing in Eq.~(\ref{fr-1}). It means that the arbitrary transfer matrix
corresponding to the Young tableau with more than one column can be expressed it terms of
the "one column" transfer matrices. Such expression was obtained by Bazhanov and
Reshetikhin~\cite{Bazhanov:1989yk}
\begin{eqnarray}\label{TL}
 \bar t_\ell(u)=\det A(u)\,, && A_{ij}(u)=\bar t_{m_j-j+i}(u+i-1),  1\leq i,j\leq \ell_1\,.
\end{eqnarray}
Here $\bar t_{\bel}(u)\equiv t_\bel(u+f_\bel)/t_0(u)$, $m_j$ is the length of the $j-$th
column in the Young tableau $\bel$. The "one-column" transfer matrices $\bar t_m(u)$ are
defined in~(\ref{btk}) for $0\leq m\leq N$ and $\bar t_m(u)\equiv 0$ for $m<0$ and $m>N$.
To verify~(\ref{TL}) it is sufficient to show that it solves Eq.~(\ref{fr-1}) that can be
done with the help of the identity~(\ref{ab}).

%%%%%%%%%%%%%%%%%%%%%%%%%%%%%%%%%%%%%%%%%%%%%%%%%%%%%%%%%%%%%%%%%%%%%%%%%%%%%%%%%%%%%%%%%%
\section{Summary}
%%%%%%%%%%%%%%%%%%%%%%%%%%%%%%%%%%%%%%%%%%%%%%%%%%%%%%%%%%%%%%%%%%%%%%%%%%%%%%%%%%%%%%%%%%
We study the properties of transfer matrices for the $sl(N)$ spin chain models. It was
shown in Ref.~\cite{slN} that the transfer matrices with a generic (infinite-dimensional)
auxiliary space are factorized into the product of $N-$commuting Baxter
$\mathcal{Q}-$operators. Both transfer matrices and Baxter operators depend on the
regularization parameter $\tau$ which  ensures the convergence of the corresponding
traces. The regularized transfer matrices are invariant only with respect to the Cartan
generators of the $sl(N)$ algebra and become singular when the regularization is removed,
$\tau\to 1$. The same concerns the Baxter operators $\mathcal{Q}_j(u,\tau)$ which are
singular at $\tau\to 1$, except the operator $\mathcal{Q}_{N}(u,\tau)$ which is finite in
this limit. To find operators which survive removing of the regularization we considered
the transfer matrices with  the special auxiliary spaces. It was shown that the transfer
matrix with the auxiliary space $\mathbb{V}^{(k)}_\brho$~(see Eq.~(\ref{Vk})) is
factorized into the product of $N-k$ commuting operators: $\mathcal{Q}_{j}(u,\tau)$,
$j=1,\ldots,N-k-1$ and one new operator which is finite in the limit $\tau\to 1$.
Moreover, new operator can be represented as the alternating sum over the product of the
Baxter operators~$\mathcal{Q}_j(u,\tau)$, $j=N-k,\ldots, N$. This representation follows
from the BGG-resolution for the finite dimensional $sl(N)$ modules and the factorized
representation for the generic transfer matrices.

We defined new set of the commuting operators $\widehat {\mathcal{Q}}_{j}(u,\tau)$, $j=2,\ldots,N$
which  are finite in the limit $\tau\to 1$. The finite dimensional transfer matrices  can
be expressed as the sum over  ratios of $\widehat {\mathcal{Q}}_{j}(u,\tau)$ operators. The
corresponding expressions have the same form as  in the Nested Bethe Ansatz approach.

%%%%%%%%%%%%%%%%%%%%%%%%%%%%%%%%%%%%%%%%%%%%%%%%%%%%%%%%%%%%%%%%%%%%%%%%%%%%%%%%5
\section*{Acknowledgment}

This work was supported by the RFFI grants 07-02-92166, 09-01-93108 (S.D. \& A.M.), RFFI
grants 08-01-00638, 09-01-12150 (S.D.) and by the German Research Foundation: DFG grant
9209282 (A.M.) and DFG grant GZ KI 623/7-1(S.D.).

%%%%%%%%%%%%%%%%%%%%%%%%%%%%%%%%%%%%%%%%%%%%%%%%%%%%%%%%%%%%%%%%%%%%%%%%%%%%%%%%%%%%%%%%%%%%
%\appendix
\renewcommand{\theequation}{\Alph{section}.\arabic{equation}}
\setcounter{table}{0}
\renewcommand{\thetable}{\Alph{table}}
\setcounter{section}{0}
%%%%%%%%%%%%%%%%%%%%%%%%%%%%%%%%%%%%%%%%%%%%%%%%%%%%%%%%%%%%%%%%%%%%%%%%%%%%%%%%%%%%%%%%%%
\section{Appendix: Determinant identity}
\label{nested}
%%%%%%%%%%%%%%%%%%%%%%%%%%%%%%%%%%%%%%%%%%%%%%%%%%%%%%%%%%%%%%%%%%%%%%%%%%%%%%%%%%%%%%%%%%
Let
$A_k(j_1,\ldots,j_k)$ be a determinant of  $k\times k$ matrix
\begin{eqnarray}
A_k(j_1,\ldots,j_k)=
\det\left |
\begin{array}{ccc}a_{1}(j_1)& \ldots& a_{1}(j_k)\\
                 \vdots&\ddots&\vdots\\
a_{k}(j_1)& \ldots& a_k(j_k)
\end{array}\right  |\,,
\end{eqnarray}
% \widehat{Q}_{N-1}(u)
and similarly  $A_{k-1}(j_2,\ldots,j_k)$ is the determinant of $(k-1)\times(k-1)$
matrix $a_{ik}=a_{i}(j_k)$, $2\leq i,j\leq k$.
Then the following identity holds
\begin{eqnarray}\label{Ai}
\frac{A_{k}(j_0,j_2,\ldots,j_{k})}{A_{k}(j_1,j_2,\ldots,j_{k})}&=&
\frac{A_{k-1}(j_2,\ldots,j_{k})}{A_{k-1}(j_1,\ldots,j_{k-1})}
\nonumber\\
&&+
\frac{A_{k-1}(j_0,j_2\ldots,j_{k-1})}{A_{k-1}(j_1,\ldots,j_{k-1})}
\frac{A_{k}(j_0,j_1,\ldots,j_{k-1})}{A_{k}(j_1,j_2,\ldots,j_{k})}
\,.
\end{eqnarray}
It easily follows from the identity~(\ref{ab}).


\begin{thebibliography}{99}


\bibitem{FST}
  L.~D.~Faddeev, E.~K.~Sklyanin and L.~A.~Takhtajan,
  {\it The Quantum Inverse Problem Method.~1},
  Theor.\ Math.\ Phys.\  {\bf 40} (1980) 688
  [Teor.\ Mat.\ Fiz.\  {\bf 40} (1979) 194].
%%CITATION = TMPHA,40,688;%%

\bibitem{KuSk1} P.P. Kulish and E.K. Sklyanin ,~{\it Quantum spectral transform method.
    Recent developments}, Lect. Notes in
Physics, {\bf v 151}, (1982) , 61-119.

\bibitem{Skl} E.K.Sklyanin,{\em Quantum Inverse Scattering Method.Selected Topics}, in
    "Quantum Group and Quantum Integrable Systems" (Nankai Lectures in Mathematical
    Physics), ed. Mo-Lin Ge,Singapore:World Scientific,1992,pp.63-97; hep-th/9211111

\bibitem{Fad} L.D. Faddeev,~{\it How Algebraic Bethe Ansatz works for integrable model},
    In: Quantum Symmetries/Symetries Qantiques, Proc.Les-Houches summer school, LXIV. Eds.
    A.Connes,K.Kawedzki, J.Zinn-Justin. North-Holland, 1998, 149-211, hep-th/9605187,


\bibitem{Fusion}
  P.~P.~Kulish, N.~Y.~Reshetikhin and E.~K.~Sklyanin,
  {\it Yang-Baxter Equation And Representation Theory. 1,}
  Lett.\ Math.\ Phys.\  {\bf 5} (1981) 393.
  %%CITATION = LMPHD,5,393;%%

\bibitem{KR82} P. P. Kulish\ and\ N. Yu. Reshetikhin, {\it On ${\rm GL}\sb{3}$-invariant
    solutions of the Yang-Baxter equation and associated quantum systems,} Zap. Nauchn.
    Sem. Leningrad. Otdel. Mat. Inst. Steklov. (LOMI) {\bf 120} (1982), 92--121.

\bibitem{Kuniba} A.Kuniba, T.Nakanishi, J.Suzuki,
 {\it Functional relations in
solvable lattice models I. Functional relations and representations theory II.
Applications} Int.J.Mod.Phys. {\bf A9} 1994,  5215-5312.

\bibitem{Zabrodin}
  A.~Zabrodin,
  {\it Discrete Hirota's equation in quantum integrable models,}
  arXiv:hep-th/9610039.
  %%CITATION = HEP-TH/9610039;%%

\bibitem{Tsuboi}
  Z.~Tsuboi,
  {\it Analytic Bethe ansatz and functional equations for Lie superalgebra\\
  $sl(r+1|s+1)$},
  J.\ Phys.\ A  {\bf 30} (1997) 7975.
  %%CITATION = JPAGB,A30,7975;%%

\bibitem{KLWZ} I.Krichever, O.Lipan, P.Wiegman, A.Zabrodin, {\it Quantum Integrable
    Systems
    and Elliptic Solutions of Classical Discrete Nonlinear Equations}
    arXiv:hep-th/9604080.







\bibitem{Tsuboi98}
  Z.~Tsuboi,
  {\it Analytic Bethe ansatz related to a one-parameter family of
  finite-dimensional representations of the Lie superalgebra $sl(r+1|s+1)$},
  J.\ Phys.\ A  {\bf 31} (1998) 5485.
  %%CITATION = JPAGB,A31,5485;%%




\bibitem{KR} A.~Kirillov and N.~Reshetikhin, {\it Proc. Conf. on Infinite Dimensional Lie
    Groups and Algebras}, Marseille 1988, ed. V. G. Kac, (Singapore: World Scientific)

\bibitem{Bazhanov:1989yk}
  V.~Bazhanov and N.~Reshetikhin,
  {\it Restricted Solid On Solid Models Connected With Simply Based Algebras And
  Conformal Field Theory},
  J.\ Phys.\ A  {\bf 23} (1990) 1477.
  %%CITATION = JPAGB,A23,1477;%%

\bibitem{slN}
  S.~E.~Derkachov and A.~N.~Manashov,
  {\it Factorization of $R-$matrix and Baxter $Q-$operators for generic $sl(N)$ spin
  chains},
  J.\ Phys.\ A  {\bf 42} (2009) 075204.
  %[arXiv:0809.2050 [nlin.SI]].
  %%CITATION = JPAGB,A42,075204;%%



\bibitem{SD}
  S.~E.~Derkachov,
  {\it Factorization of the $R-$matrix. I,II} , Zapiski
nauchnuch seminarov POMI, 335, p.134, [arXiv:math/0503396].
%%CITATION = MATH-QA 0503396;%%

\bibitem{DM06}
  S.~E.~Derkachov and A.~N.~Manashov,
  {\it $R-$Matrix and Baxter $Q-$Operators for the Noncompact $SL(N,C)$ Invarianit Spin
  Chain},
  SIGMA {\bf 2} (2006) 084.
 % [arXiv:nlin/0612003].
  %%CITATION = 00480,2,084;%%


\bibitem{BGG} Bernstein I.N., Gelfand I.M., Gelfand S.I., {\it Differential Operators on
    the Base Affine Space and a Study of $g-$Modules, Lie Groups and Their Representations}
    I. M. Gelfand, Ed., Adam Hilger, London, 1975.



\bibitem{BLZ-III}
  V.~V.~Bazhanov, S.~L.~Lukyanov and A.~B.~Zamolodchikov,
{\it Integrable structure of conformal field theory. III: The Yang-Baxter
  relation},
  Commun.\ Math.\ Phys.\  {\bf 200} (1999) 297.
  %[arXiv:hep-th/9805008].
%%CITATION = HEP-TH 9805008;%%

%\cite{Pronko:1999gh}
\bibitem{Pronko:1999gh}
  G.~P.~Pronko and Yu.~G.~Stroganov,
{\it   The complex of solutions of the nested Bethe ansatz: The $A(2)$ spin
  chain},
  arXiv:hep-th/9902085.
  %%CITATION = HEP-TH/9902085;%%

\bibitem{Baxter:1972hz}
  R.~J.~Baxter,
  {\it Partition function of the eight-vertex lattice model},
  Annals Phys.\  {\bf 70} (1972) 193
  [Annals Phys.\  {\bf 281} (2000) 187].
  %%CITATION = APNYA,281,187;%%

%\cite{Bazhanov:1989nc}
\bibitem{BzSt90}
  V.~V.~Bazhanov and Yu.~G.~Stroganov,
  {\it Chiral Potts model as a descendant of the six vertex model},
  J.\ Statist.\ Phys.\  {\bf 59} (1990) 799.
  %%CITATION = JSTPB,59,799;%%



\bibitem{GP92}
  M.~Gaudin and V.~Pasquier,
  {\it The periodic Toda chain and a matrix generalization of the bessel
  function's recursion relations},
  J.\ Phys.\ A  {\bf 25} (1992) 5243.
%%CITATION = JPAGB,A25,5243;%%

\bibitem{Volkov}
  A.~Y.~Volkov,
{\it Quantum lattice KdV equation},
  Lett.\ Math.\ Phys.\  {\bf 39} (1997) 313.
%  [arXiv:hep-th/9509024].
%%CITATION = LMPHD,39,313;%%


%\cite{Derkachov:1999pz}
\bibitem{SDQ}
  S.~E.~Derkachov,
  {\it Baxter's $Q-$operator for the homogeneous $XXX$ spin chain},
  J.\ Phys.\ A~{\bf 32} (1999) 5299.
  %[arXiv:solv-int/9902015].
%%CITATION = JPAGB,A32,5299;%%

\bibitem{Hikami}
  K.~Hikami,
  {\it Baxter Equation for Quantum Discrete Boussinesq Equation},
  Nucl.\ Phys.\  B~{\bf 604} (2001) 580.
%  [arXiv:nlin/0102021].
%%CITATION = NUPHA,B604,580;%%



\bibitem{Pronko}
  G.~P.~Pronko,
  {\it On the Baxter's $Q$ operator for the $XXX$ spin chain},
  Commun.\ Math.\ Phys.\  {\bf 212} (2000) 687.
 % [arXiv:hep-th/9908179].
%%CITATION = CMPHA,212,687;%%

\bibitem{SklyaninB} E.~K.~Sklyanin, {\it B\"acklund transformations and Baxter's
    Q-operator},
    [arXiv:nlin/0009009 [nlin.SI]].

\bibitem{SL2C}
  S.~E.~Derkachov, G.~P.~Korchemsky and A.~N.~Manashov,
 {\it Noncompact Heisenberg spin magnets from high-energy QCD. I: Baxter
  $Q-$operator and separation of variables,}
  Nucl.\ Phys.\ B {\bf 617} (2001) 375.
%  [arXiv:hep-th/0107193].
%%CITATION = HEP-TH 0107193;%%

\bibitem{DKM-I}
  S.~E.~Derkachov, G.~P.~Korchemsky and A.~N.~Manashov,
  {\it Separation of variables for the quantum $SL(2,R)$ spin chain,}
  JHEP {\bf 0307} (2003) 047.
%  [arXiv:hep-th/0210216].
%%CITATION = HEP-TH 0210216;%%


\bibitem{RW}
  M.~Rossi and R.~Weston,
  {\it A Generalized Q operator for $U(q)$(affine $sl(2)$) vertex models},
  J.\ Phys.\ A  {\bf 35} (2002) 10015.
 % [arXiv:math-ph/0207004].
%%CITATION = JPAGB,A35,10015;%%

\bibitem{KMS}
  V.~B.~Kuznetsov, V.~V.~Mangazeev and E.~K.~Sklyanin,
 {\it $Q-$operator and factorised separation chain for Jack's symmetric
  polynomials},
  Indag.\ Math.\  {\bf 14} (2003) 451.
%  [arXiv:math/0306242].
  %%CITATION = IMTHB,14,451;%%

%%CITATION = MATH-CA 0306242;%%

\bibitem{DKM-II}
  S.~E.~Derkachov, G.~P.~Korchemsky and A.~N.~Manashov,
   {\it Baxter $Q-$operator and separation of variables for the open $SL(2,R)$ spin chain,}
  JHEP {\bf 0310} (2003) 053.
%  [arXiv:hep-th/0309144].
%%CITATION = HEP-TH 0309144;%%

\bibitem{Korff04} C.~Korff, {\it Auxiliary matrices for the six-vertex model and the
    algebraic
    Bethe ansatz},\
 J.~Phys.\ A  {\bf 37} (2004) 7227.
%[arXiv:math-ph/0404028]



\bibitem{Korff-c} C.~Korff, {\it Representation Theory and Baxter's $TQ-$equation for the
    six-vertex model. A pedagogical overview} [arXiv:cond-mat/0411758].


\bibitem{Bytsko}
  A.~G.~Bytsko and J.~Teschner,
 {\it Quantization of models with non-compact quantum group symmetry: Modular
  $XXZ$ magnet and lattice sinh-Gordon model,}
  J.\ Phys.\ A  {\bf 39} (2006) 12927.
%  [arXiv:hep-th/0602093].
%%CITATION = JPAGB,A39,12927;%%

\bibitem{BLZ}
  V.~V.~Bazhanov, S.~L.~Lukyanov and A.~B.~Zamolodchikov,
  {\it Integrable structure of conformal field theory, quantum KdV theory and
  thermodynamic Bethe ansatz,}
  Commun.\ Math.\ Phys.\  {\bf 177} (1996) 381.
  %[arXiv:hep-th/9412229].
  %%CITATION = CMPHA,177,381;%%





\bibitem{BLZ-II}
  V.~V.~Bazhanov, S.~L.~Lukyanov and A.~B.~Zamolodchikov,
 {\it Integrable Structure of Conformal Field Theory II. Q-operator and DDV
  equation,}
  Commun.\ Math.\ Phys.\  {\bf 190} (1997) 247.
 % [arXiv:hep-th/9604044].
%%CITATION = CMPHA,190,247;%%







\bibitem{BHK}
  V.~V.~Bazhanov, A.~N.~Hibberd and S.~M.~Khoroshkin,
{\it Integrable structure of $W(3)$ conformal field theory, quantum Boussinesq
 theory and boundary affine Toda theory},
  Nucl.\ Phys.\ B {\bf 622}, 475 (2002).
 % [arXiv:hep-th/0105177].
%%CITATION = HEP-TH 0105177;%%


\bibitem{TsuboiB}
  V.~V.~Bazhanov and Z.~Tsuboi,
  {\it Baxter's $Q-$operators for supersymmetric spin chains,}
  Nucl.\ Phys.\  B {\bf 805} (2008) 451.
%  [arXiv:0805.4274 [hep-th]].
  %%CITATION = NUPHA,B805,451;%%

%\cite{Kojima:2008zza}
\bibitem{Kojima}
  T.~Kojima,
{\it Baxter's $Q-$operator for the W-algebra $W_N$,}
  J.\ Phys.\ A  {\bf 41} (2008) 355206.
  %[arXiv:0803.3505 [Unknown]].
  %%CITATION = JPAGB,A41,355206;%%

%\cite{Bazhanov:2010ts}
\bibitem{Bazhanov:2010ts}
  V.~V.~Bazhanov, T.~Lukowski, C.~Meneghelli and M.~Staudacher,
{\it A Shortcut to the $Q-$Operator},
  arXiv:1005.3261 [hep-th].
  %%CITATION = ARXIV:1005.3261;%%

  \bibitem{DMsl2}
  S.~E.~Derkachov and A.~N.~Manashov,
 {\it Factorization of the transfer matrices for the quantum $sl(2)$ spin chains
  and Baxter equation,}
  J.\ Phys.\ A  {\bf 39} (2006) 4147.
%  [arXiv:nlin/0512047].
  %%CITATION = JPAGB,A39,4147;%%



%\cite{Derkachov:2006fz}
\bibitem{DMsl3}
  S.~E.~Derkachov and A.~N.~Manashov,
{\it Baxter operators for the quantum $sl(3)$ invariant spin chain,}
  J.\ Phys.\ A  {\bf 39} (2006) 13171.
%  [arXiv:nlin/0604018].
  %%CITATION = JPAGB,A39,13171;%%




\bibitem{BDKM-21}
  A.~V.~Belitsky, S.~E.~Derkachov, G.~P.~Korchemsky and A.~N.~Manashov,
{\it Baxter $Q-$operator for graded $SL(2|1)$ spin chain,}
  J.\ Stat.\ Mech.\  {\bf 0701} (2007) P005.
 % [arXiv:hep-th/0610332].
%%CITATION = JSTAT,0701,P005;%%

\bibitem{Verma} Verma N., {\it Structure of certain induced representations of complex
    semisimple Lie algebras}, Bull. Amer. Math. Soc. 74 (1968).



\bibitem{Zelobenko} \v{Z}elobenko, D.~P., {\it  Compact Lie Groups and Their
    Representations}, Providence, R.I., Amer.~Math.~Soc., 1973

\bibitem{Khoroshkin}
  S.~M.~Khoroshkin and V.~N.~Tolstoi,
 {\it  Yangian Double And Rational R Matrix,}
  arXiv:hep-th/9406194.
  %%CITATION = HEP-TH/9406194;%%

\bibitem{Sutherland}
  B.~Sutherland,
 {\it A General Model For Multicomponent Quantum Systems,}
  Phys.\ Rev.\  B {\bf 12} (1975) 3795.
  %%CITATION = PHRVA,B12,3795;%%

\bibitem{Kulish83}
  P.~P.~Kulish and N.~Y.~Reshetikhin,
 {\it Diagonalization Of $GL(N)$ Invariant Transfer Matrices And Quantum N Wave
  System (Lee Model),}
  J.\ Phys.\ A  {\bf 16} (1983) L591.
%%CITATION = JPAGB,A16,L591;%%


\end{thebibliography}
\end{document}